\documentclass[jcp,aip,amsmath,floatfix,reprint]{revtex4-1}

\usepackage{txfonts}
\usepackage{bm}
\usepackage{color}
\usepackage{dcolumn}
\usepackage{graphicx}

\begin{document}
\title{Active space decomposition with multiple sites: Density matrix renormalization group algorithm}
\author{Shane M. Parker}
\author{Toru Shiozaki}
\affiliation{Department of Chemistry, Northwestern University, 2145 Sheridan Rd., Evanston, IL 60208, USA.}
\date{\today}
\begin{abstract}
We extend the active space decomposition method, recently developed by us, to more than two active sites using
the density matrix renormalization group algorithm. The fragment wave functions are described by complete or restricted active-space wave functions.
Numerical results are shown on a benzene pentamer and a perylene diimide trimer.
It is found that the truncation errors in our method decrease almost exponentially with respect to the number of renormalization states $M$,
allowing for numerically exact calculations (to a few $\mu E_\mathrm{h}$ or less) with $M = 128$ in both cases.
This rapid convergence is because the renormalization steps are used only for the interfragment electron correlation.
\end{abstract}
\maketitle

The electronic structure of organic aggregates and crystals has been studied to characterize and predict electron and exciton dynamics in
organic photovoltaics and light emitting devices.\cite{Coropceanu2007CR}
While the majority of computational studies of such processes has been performed using density functional theory (DFT), processes that involve
multiple excitons, such as singlet fission,\cite{Smith2010CR} cannot be well described by standard time-dependent DFT, which has stimulated the development of
accurate wave function theories for these systems.

We have recently developed a method, called active-space decomposition (ASD),\cite{Parker2013JCP} which alleviates the large computational
cost of active-space methods by tailoring the wave function ansatz to molecular dimers. In ASD, a dimer wave function $\Psi$ is
compactly expressed as a linear combination of tensor products of monomer wave functions,
i.e.,
\begin{equation}
  |\Psi\rangle = \sum_{IJ}C_{IJ}|\Phi^A_I\rangle \otimes |\Phi^B_J\rangle \equiv \sum_{IJ} C_{IJ} |\Phi^A_I\Phi^B_J\rangle,
\end{equation}
where $A$ and $B$ label monomers, and $I$ and $J$ label orthonormal monomer states.
This expression is exact when the complete set of states is included in the sum; practically the number of states in the summation is set to a finite value.
We have shown\cite{Parker2013JCP} that by choosing monomer wave functions that diagonalize a monomer Hamiltonian within the orthogonal subspace
(i.e., $\hat{H}^A|\Phi^A_I\rangle = E^A_I|\Phi^A_I\rangle$),
dimer wave functions rapidly converge to the exact solution with respect to the number and type of monomer states.
This product basis is also related to quasi-diabatic representation of the model space.\cite{Parker2014JCTC}
ASD is compatible with any wave function method
for which transition density matrices are available. In particular, efficient implementation of ASD has been realized with
complete active space (CAS) and restricted active space\cite{Olsen1988JCP} (RAS) monomer wave functions.
ASD has been used to compute model Hamiltonians for
dynamical processes such as singlet fission in tetracene and pentacene dimers\cite{Parker2014JPCC} and charge and
triplet energy transfer in benzene dimers.\cite{Parker2014JCTC}

Here, we extend ASD beyond dimers to incorporate an arbitrary number of fragments:
\begin{equation}
  |\Psi\rangle = \sum_{IJK\cdots}C_{IJK\cdots}|\Phi^A_I\rangle \otimes |\Phi^B_J\rangle \otimes |\Phi^C_K\rangle \otimes \cdots,
\end{equation}
where $I$, $J$, and $K$ formally run over the complete set of states on each fragment.
The coefficient tensor $C_{IJK\cdots}$ is approximated by matrix product states used in the density
matrix renormalization group (DMRG) algorithm,
\begin{equation}
  C_{IJK\cdots} = \mathrm{Tr}[\mathbf{C}^{A,I}\mathbf{C}^{B,J}\mathbf{C}^{C,K}\cdots], \label{cmps}
\end{equation}
in which the dimension of matrices $\mathbf{C}$ is set to a predetermined value $M$.
First described by White and used to compute ground
and excited states of the Heisenberg spin chain,\cite{White1992PRL,White1993PRB} DMRG has proven to
be a powerful tool that provides numerically exact solutions with polynomial cost even for
strongly correlated systems.\cite{White1999JCP,Zgid2009ARCC,Kurashige2009JCP,Marti2010ZPC,Chan2011AnnRevPhys,Wouters2014EPJD}
In chemical applications, DMRG has enabled highly accurate calculations on strongly correlated electronic structure of molecules or complexes
that are intractable with traditional active-space methods.\cite{Kurashige2013NC,Sharma2014NC}

\begin{figure}[t]
  \includegraphics[width=0.50\textwidth]{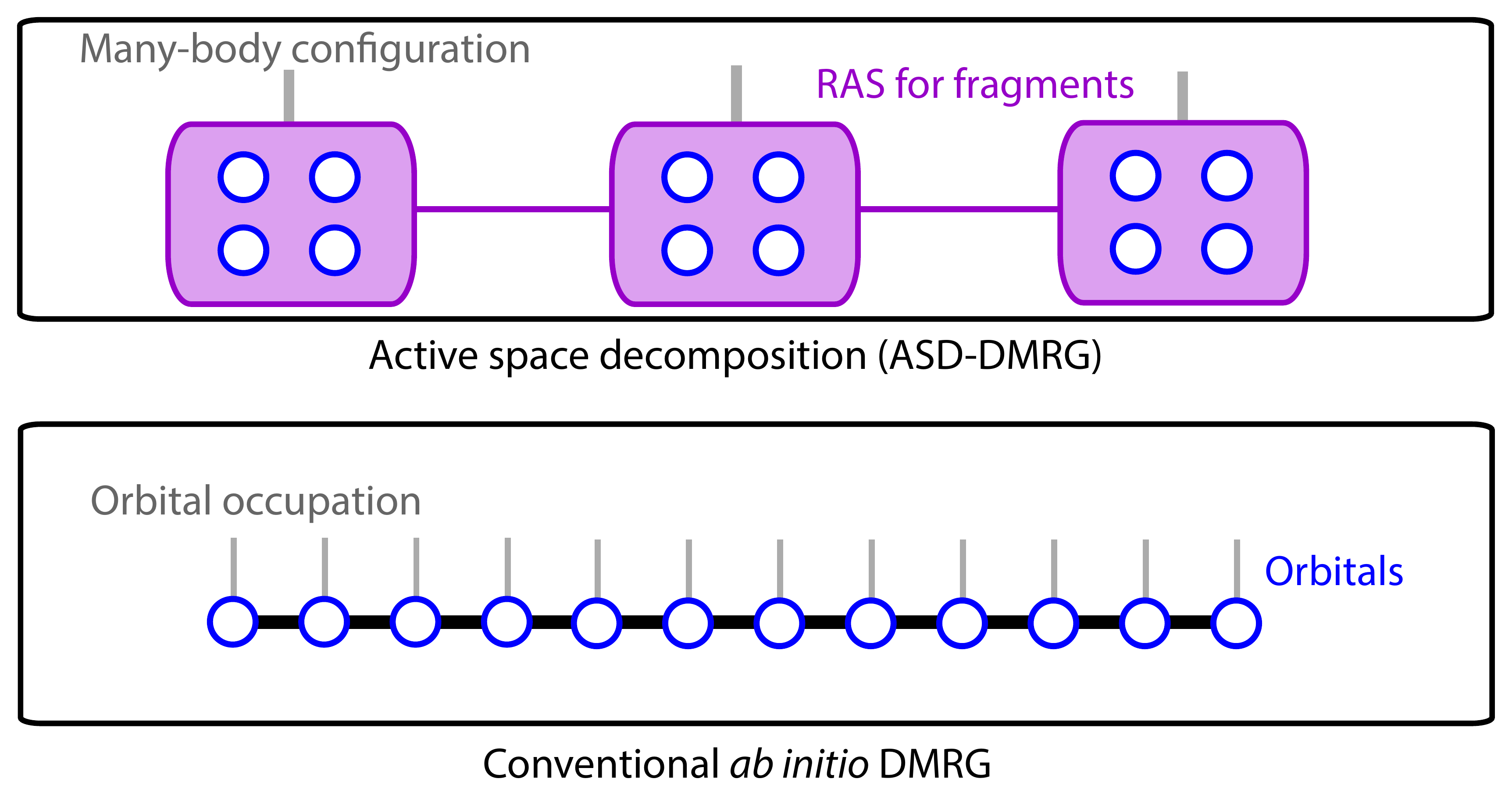}
  \caption{\label{fig:mps}Comparison of wave functions in ASD-DMRG and in conventional {\it ab initio} DMRG.}
\end{figure}

We start by sketching the ASD-DMRG algorithm,
which is formulated in terms of a one-dimensional chain of $n$ sites (see Fig.~\ref{fig:mps}).
In the conventional {\it ab initio} DMRG, these sites are chosen to be one-electron orbitals.\cite{White1999JCP}
In this work, we choose each site to be the CAS or RAS wave function of a single molecule or fragment,
alleviating the orbital ordering problem in the conventional {\it ab initio} DMRG algorithm\cite{Moritz2005JCP,Rissler2006CP}
and allowing for applications to three-dimensional structures.
As a consequence, the dimensionality of the single site, $d$, is much larger in ASD-DMRG than in conventional approaches.
However, a key result of this study is that this allows $M$ to be very small. 
Since calculations involving the full configurational
space of a molecular dimer are intractable for large molecules of materials interest, we have implemented the
single-site variant of the DMRG algorithm.\cite{White2005PRB}
The single-site algorithm converges more smoothly than the two-site algorithm,
though it is more prone to getting stuck in local minima during the optimization (see below).\cite{White2005PRB, Zgid2008JCP2}

The chain is initiated by computing
a user-defined number of low-lying eigenstates of the first site. We will represent the $i$-th single site as $\bullet_i$.
These states are then renormalized to form a block, i.e. $L_1\leftarrow\bullet_1$ with $L_i$ referring to a block
containing $i$ sites.
Next, a new site is added to the chain to form $L_1\bullet_2$, and several low-lying eigenstates of this block-site
system are computed and renormalized as $L_2\leftarrow L_1\bullet_2$. This process is continued until
each site has been added to the chain. During this growing phase, the sites that have yet to be incorporated
into the block are included at the mean-field level. This is equivalent to using a tensor product of the Hartree--Fock
ground state of each site as the initial guess. Such a mean-field initial guess is well-defined in the case where each site is a distinct molecule, although other initial guesses may be necessary in the case where sites
are covalently bound.

The final stage of the algorithm involves ``sweeping'' along
the chain to optimize the renormalized states. In each step of the sweep,
a system $L_{i-1}\bullet_i R_{n-i}$ is diagonalized, which contains a left block describing $i-1$ sites $L_{i-1}$,
a right block describing
$n-i$ sites $R_{n-i}$, and the $i$-th single site, $\bullet_i$.
The wave function at this step is written as
\begin{align}
  |\Psi\rangle &= \sum^M_{\mu\nu} \sum_\phi C^{i,\phi}_{\mu\nu}|\Phi^{L_{i-1}}_\mu \rangle \otimes |\phi^i\rangle \otimes |\Phi^{R_{n-i}}_\nu\rangle \nonumber\\
               &= \sum^M_{\mu\nu}|\Phi^{L_{i-1}}_\mu \Phi^i_{\mu\nu} \Phi^{R_{n-i}}_\nu\rangle,
\end{align}
where $|\phi^i\rangle$ is a Slater determinant only with orbitals on the site $i$, $|\Phi^{L_{i-1}}_\mu\rangle$ is
a renormalized left-block state, $|\Phi^{R_{n-i}}_\nu\rangle$ is a renormalized right-block state, and
$|\Phi^i_{\mu\nu}\rangle\equiv\sum_\phi C^{i,\phi}_{\mu\nu} |\phi\rangle$.
The key insight behind the DMRG algorithm is that
the optimal block states to represent the combined $L_{i-1}\bullet_i$ part of the chain can be obtained by
the Schmidt decomposition of the reduced density matrix (RDM),
\begin{equation} \label{eq:rdm}
  \hat{\rho}_{L_{i-1}\bullet_i} = \text{Tr}\left[ |\Psi\rangle\langle\Psi|\right]_{R_{n-i}},
\end{equation}
where the trace is over all states in
the right block.
Thus, one computes and diagonalizes the RDM in each step of the sweep and uses the $M$ eigenvectors
corresponding to the largest eigenvalues to renormalize $L_{i}\leftarrow L_{i-1}\bullet_i$.

To diagonalize the Hamiltonian during the sweep, we resolve
the Hamiltonian in terms of the renormalized block states $\{\Phi^L_\mu\}$ and $\{\Phi^R_\nu\}$.
Defining $|\Phi^B_{\mu\nu}\rangle\equiv|\Phi^L_\mu\rangle\otimes|\Phi^R_\nu\rangle$ and following the notation of Kurashige and
Yanai,\cite{Kurashige2009JCP} we write the Hamiltonian as
\begin{align} \label{eq:ham}
  \hat{H}= &\hat{H}^B + \hat{H}^i
  +   \sum_{p_B}\left(\hat{T}^i_{p_B}\hat{p}_B + \hat{p}_B^\dagger\hat{T}^{i\dagger}_{p_B} \right) \nonumber\\
  + & \sum_{p_i}\left(\hat{S}^B_{p_i}\hat{p}_i  + \hat{p}_i^\dagger\hat{S}^{B\dagger}_{p_i} \right)
  +   \sum_{p_iq_i}\hat{p}_i^\dagger\hat{q}_i\hat{Q}^B_{p_iq_i} \nonumber\\
  + & \sum_{p_ir_i}\left(\hat{p}_i^\dagger\hat{r}_i^\dagger\hat{P}^B_{p_ir_i} + \hat{P}^{B\dagger}_{p_ir_i}\hat{p}_i\hat{r}_i \right),
\end{align}
in which we have introduced the following operators:
\begin{align}
  \hat{H}^X =& \sum_{p_Xq_X}\hat{p}_X^\dagger\hat{q}_X h_{p_Xq_X} + \frac{1}{2} \sum_{p_Xq_Xr_Xs_X}\hat{p}_X^\dagger \hat{q}_X^\dagger \hat{r}_X \hat{s}_X (p_Xs_X|q_Xr_X),\\
  \hat{T}^i_{p_B} =& \sum_{q_ir_is_i}\hat{s}_i^\dagger\hat{r}_i^\dagger \hat{q}_i (p_Bs_i|q_ir_i), \\
  \hat{S}^B_{p_i} =& \sum_{q_B} \hat{q}_B^\dagger h_{p_iq_B} + \sum_{q_B,r_B,s_B}\hat{s}_B^\dagger\hat{r}_B^\dagger \hat{q}_B (p_is_B|q_Br_X), \\
  \hat{P}^B_{p_ir_i} =& \sum_{q_is_i}\hat{q}_i\hat{s}_i (p_Bs_i|q_ir_B),\\
  \hat{Q}^B_{p_iq_i} =& \sum_{r_is_i}\hat{r}_i^\dagger\hat{s}_i \left[ (p_Bq_B|r_is_i) - (p_Br_i|q_Bs_i) \right].
\end{align}
Here $X$ is either $B$ or $i$, $\hat{p}_X$ ($\hat{p}_X^{\dagger})$ is the annihilation (creation) operator
for orbital $p$ on site/block $X$, $h_{pq}$ is a matrix element of the one-electron operator and $(pq|rs)$ is a matrix
element of the two-electron Coulomb operator.
In our algorithm, we compute and store transition density matrices whose indices belong to the same block using an algorithm developed in Refs.~\onlinecite{Parker2013JCP,Parker2014JCTC}.
For instance,
\begin{align}
\Gamma^{Z,\lambda'\lambda}_{p^\dagger q} \equiv \langle \Phi^Z_{\lambda'}|\hat{p}^\dagger\hat{q}|\Phi^Z_\lambda\rangle,
\end{align}
where $Z$ is either $L$ or $R$, $\lambda$ is either $\mu$ or $\nu$,
and $\Gamma^{Z,\lambda'\lambda}_{p^\dagger q^\dagger r}$, $\Gamma^{Z,\lambda'\lambda}_{p q}$, and $\Gamma^{Z,\lambda'\lambda}_{p}$ are likewise defined.
The block operators of Eq.~\eqref{eq:ham} are resolved using these transition densities, e.g.,
\begin{align}
  \langle \Phi^B_{\mu'\nu'}|\hat{P}^B_{p_ir_i}&|\Phi^B_{\mu\nu}\rangle =
      \delta_{\nu'\nu}\sum_{q_Ls_L}\Gamma^{L,\mu'\mu}_{q_Ls_L} (p_is_L|r_iq_L) \nonumber\\
    + & \delta_{\mu'\mu}\sum_{q_Rs_R}\Gamma^{R,\nu'\nu}_{q_Rs_R} (p_is_R|r_iq_R) \nonumber\\
    - & (-1)^{N_{L_\mu}}\sum_{q_R}\Gamma^{R,\nu'\nu}_{q_R}\sum_{s_L}\Gamma^{L,\mu'\mu}_{s_L} (p_is_L|r_iq_R) \nonumber\\
    + & (-1)^{N_{L_\mu}}\sum_{q_L}\Gamma^{L,\mu'\mu}_{q_L}\sum_{s_R}\Gamma^{R,\nu'\nu}_{s_R} (p_is_R|r_iq_L),
\end{align}
where $N_{L_\mu}$ is the number of electrons in $L_\mu$.

Active orbitals are identified as follows. We start by picking orbitals from
monomer Hartree--Fock calculations using minimal basis. Then we perform Hartree--Fock on the full system and localize the canonical
molecular orbitals to fragments using the Pipek--Mezey algorithm.\cite{Pipek1989JCP} Finally, overlaps between
these fragment-localized orbitals and the minimal-basis orbitals are used to automatically select
active orbitals on each site.

\begin{table*}[t]
  \caption{\label{table:benzene}
    Total energies computed using $M=128$ and
    energy differences obtained with selected smaller values of $M$
    for stacks of $n$ benzene molecules (in $E_\mathrm{h}$).
  }
  \begin{ruledtabular}
  \begin{tabular}{lcdddddd}
       &   & \multicolumn{5}{c}{$\Delta E$} & \multicolumn{1}{c}{$E_\text{tot}$} \\
    \multicolumn{1}{c}{$n$} & \multicolumn{1}{c}{Active space} & \multicolumn{1}{c}{$M=1$} & \multicolumn{1}{c}{$M=4$} & \multicolumn{1}{c}{$M=8$} & \multicolumn{1}{c}{$M=16$} & \multicolumn{1}{c}{$M=32$} & \multicolumn{1}{c}{$M=128$} \\ \hline
    2 & CAS(12,12) & 0.000501 & 0.000168 & 0.000134 & 0.000069 & < 0.000001 & -461.179327 \\
    3 & CAS(18,18) & 0.001018 & 0.000343 & 0.000274 & 0.000141 & 0.000001 & -691.768264 \\
    4 & CAS(24,24) & 0.001537 & 0.000520 & 0.000415 & 0.000214 & 0.000002 &  -922.357173 \\
    5 & CAS(30,30) & 0.002055 & 0.000696 & 0.000557 & 0.000287 & 0.000002 & -1152.946072 \\
  \end{tabular}
  \end{ruledtabular}
\end{table*}

As is commonly known,\cite{White2005PRB, Zgid2008JCP2} the one-site DMRG algorithm is susceptible to getting stuck
in local minima of the parameter space due to the fact that the algorithm does not spontaneously expand
the range of quantum numbers in each site or block. For example, if a poor initial guess for the chain
includes only neutral fragments and the total charge is constrained to be neutral, then the algorithm
will keep only neutral fragment states although charge transfer configurations may be important
in the exact ground state.  To remedy this problem, we use the
perturbative correction of White\cite{White2005PRB} to the RDM [Eq.~\eqref{eq:rdm}], in which the RDM
is supplemented by
\begin{align}
  \hat{\rho}_{L_{i-1}\bullet_i}' = \hat{\rho}_{L_{i-1}\bullet_i} +
    a \sum_\tau \hat{O}^{i}_{\tau} \hat{\rho}_{L_{i-1}\bullet_i} \hat{O}^{i\dagger}_{\tau},
\end{align}
where $a>0$ is an arbitrary constant, and $\hat O^i_\tau$ is a site Hamiltonian operator
(we use $\hat{p}_i$, $\hat{p}_i^\dagger$ and $\hat{p}_i^\dagger\hat{q}_i$). The constant $a$ is used
to control the strength of the perturbation during the sweep. We typically start with a value of
$10^{-3}$ and decrease it by one order of magnitude when the energy from successive sweeps changes by less than
$10^{-7}$~$E_\mathrm{h}$. When $a$ reaches a minimum value ($a < 10^{-7}$ in our calculations) the perturbation is turned off.
With this scheme, we have seen that the final converged energies have no dependence on the initial guess.

\begin{figure}
\includegraphics[width=0.40\textwidth]{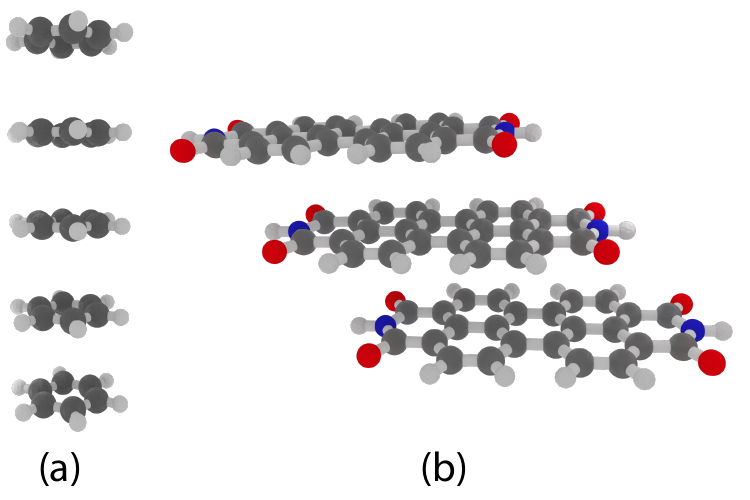}
\caption{\label{fig:stacks}Geometries of the (a) benzene and (b) PDI stacks used in this work.}
\end{figure}

Next, we show the numerical results on the ground state of a benzene stack consisting of 2 to 5
molecules with the def2-SVP basis set.\cite{ Weigend2005PCCP} The molecules are arranged cofacially with a separation of 4.0~{\AA} (see Fig.~\ref{fig:stacks}a). The geometry of each benzene molecule was obtained
in $D_{6{\text{h}}}$ symmetry using DFT with B3LYP functional\cite{Becke1993JCP,Lee1988PRB} and
the def2-TZVPP basis set.\cite{Weigend2005PCCP}
Geometry optimization was performed using {\sc turbomole}.\cite{Furche2014WIREs} Each benzene
molecule is a single site containing six $\pi$ orbitals and six electrons. Using CAS wave functions on each site,
the full stack is the equivalent of CAS(6$n$,6$n$). Calculations on a benzene dimer [for which the configuration interaction in CAS(12,12) is
tractable] and those on a benzene trimer with a smaller active space\cite{supp}
indicate that ASD-DMRG with $M=128$ recovered the CAS configuration interaction to within $10^{-7}$
$E_\mathrm{h}$. Thus, we consider $M=128$ to be fully converged for all benzene stack calculations.
In Table \ref{table:benzene}, the total energies are compared to those obtained using smaller
values of $M$.
It is worth noting that the errors decrease almost exponentially with respect to $M$,
and with as small as $M=32$, numerically exact results were obtained.
The pentamer calculation with $M=128$ required about 2200 seconds per sweep on 128 CPU cores (9 sweeps) whereas the corresponding CAS(30,30)
calculation, with $2.4\times10^{16}$ total configurations, is intractable even for today's most
powerful supercomputers.

One of the advantages of ASD-DMRG over conventional DMRG is that occupation restrictions can be introduced on individual sites.
To show this, we computed the ground state of the perylene diimide (PDI)
dimer and trimer using RAS wave functions for each molecule (def2-SVP). PDI crystallizes into a slip-stacked
configuration as shown in Fig.~\ref{fig:stacks}b.\cite{Guillermet2006TSF} For monomers, we used geometries optimized using DFT
under $D_{2\text{h}}$ symmetry with the def2-TZVPP basis set\cite{Weigend2005PCCP} and the B3LYP density functional.\cite{Becke1993JCP,Lee1988PRB}
The dimer and trimer geometries were generated
by adding a constant offset to the base unit of $(-3.34, 1.11, 3.38)$~{\AA} where $x$ is the long axis of the molecule,
$y$ is the short axis of the molecule, and $z$ is normal to the plane of the molecule. For single-site wave
functions, we use RAS with 7 orbitals in RAS~I, 4 orbitals in RAS~III, and 7 orbitals in RAS~II and a maximum of one
hole in RAS~I and one particle in RAS~III. The tensor product of multiple RAS wave functions corresponds to a more
general set of occupation restrictions than RAS itself: Each site can have one hole (particle) in RAS~I (RAS~III)
rather than there being one hole (particle) allowed for the whole stack. The dimensionality of the product wave
functions for the dimer and trimer is about $1\times10^7$ and $2\times10^{12}$, respectively.

The total energies computed for the PDI dimer and trimer are shown in Table~\ref{table:pdi}. For the PDI dimer, we
used up to $M=128$. As in the benzene case, the convergence of the total energy with respect to $M$ appears to
be exponential. On the basis of these calculations we chose $M=48$ to compute the ground state of the PDI trimer.
The energy difference between $M=128$ and $M=48$ for the dimer is about 0.06~$\mathrm{m}E_\mathrm{h}$.
Noting that error in the total energy of the benzene stacks for a fixed $M$ was roughly proportional to $n-1$,
we expect the total energy for the PDI trimer at $M=48$ to be accurate to within
0.2~$\mathrm{m}E_\mathrm{h}$. The trimer calculation with $M=48$ required ca.~1500 seconds per sweep using 256 CPU
cores,
although our program can be further optimized.

\begin{table}[t]
  \caption{\label{table:pdi}
    Total energies (in $E_\mathrm{h}$) obtained for the PDI dimer and trimer at selected values of $M$.
  }
  \begin{ruledtabular}
  \begin{tabular}{rdd}
    \multicolumn{1}{c}{$M$} & \multicolumn{1}{c}{dimer} & \multicolumn{1}{c}{trimer} \\ \hline
    1 & -2644.328613 & -3966.487183 \\
    4 & -2644.331085 & -3966.492073 \\
    8 & -2644.331681 & -3966.493276 \\
    16 & -2644.331990 & -3966.493900 \\
    32 & -2644.332199 & -3966.494332 \\
    48 & -2644.332283 & -3966.494512 \\
    72 & -2644.332327 & \multicolumn{1}{c}{\text{---}} \\
    96 & -2644.332339 & \multicolumn{1}{c}{\text{---}} \\
    128 & -2644.332343 & \multicolumn{1}{c}{\text{---}} \\
  \end{tabular}
  \end{ruledtabular}
\end{table}

In conclusion, we have generalized the ASD method to incorporate multiple fragments (i.e., active sites).
In this work the electronic structure of each fragment has been described by CAS and RAS wave functions.
Our method is based on the matrix-product approximation of the coefficients [Eq.~(\ref{cmps})], in which the wave functions
are optimized using a density matrix renormalization group algorithm.
We have shown that results with chemical accuracy (i.e., 1~$\mathrm{m}E_\mathrm{h}$ for total energies) can be obtained with a fairly small number of renormalization states $M$.
This rapid convergence of the ASD-DMRG energy with respect to $M$ is the result of breaking the system
into physically motivated fragments such that the renormalization steps are used to describe the
interchromophore electron correlation, and not the intrachromophore correlation.
Spin adaptation\cite{Zgid2008JCP1,Sharma2012JCP} has not been done in this work.
Application of ASD-DMRG to covalently bonded systems and investigation of the effects of site ordering will be studied in the future.
All the programs reported in this work have been implemented in the open-source {\sc bagel} package.\cite{bagel}

This work has been supported by Office of Basic Energy Sciences, U.S. Department of Energy (Grant No. DE-FG02-13ER16398).

\bibliography{dmrg}
\end{document}